\documentclass[lettersize,journal]{IEEEtran}
\usepackage{amsmath,amsfonts}
\usepackage{algorithmic}
\usepackage{algorithm}
\usepackage{array}
\usepackage[caption=false,font=normalsize,labelfont=sf,textfont=sf]{subfig}
\usepackage{textcomp}
\usepackage{stfloats}
\usepackage{url}
\usepackage{verbatim}
\usepackage{graphicx}
\usepackage{cite}
\usepackage{hyperref, color, amsmath, graphicx}
\usepackage[utf8]{inputenc}
\usepackage{multirow}
\usepackage{booktabs}
\usepackage{array}
\usepackage{adjustbox}
\usepackage{siunitx}
\usepackage{pifont}

\begin{document}

\newcommand{\arthur}[1]{\textcolor{cyan}{#1}}

\title{EG-VAE: A Unified Framework for Electric Guitar \\ Tone Transfer and Removal}

\author{Yen-Tung Yeh$^{1}$, Yun-Ning (Amy) Hung$^{2}$, Yi-Hsuan Yang$^{1}$\\
\vspace{2mm}
        $^{1}$ National Taiwan University \;\;\; $^{2}$ Moises \\

\thanks{}
}



\maketitle

\begin{abstract}
Electric guitar tone transfer (EGTT) and tone removal (EGTR) are two fundamental tasks in guitar tone modeling: EGTT replaces a recording's tone with that of a reference, while EGTR recovers the dry direct-input (DI) signal from a wet, processed recording. Despite their highly related nature, prior work has addressed them independently, and both works have yet to achieve satisfactory results. In this paper, we propose EG-VAE, a unified framework that jointly models EGTT and EGTR by disentangling frame-level content and global tone representations from wet recordings with a variational autoencoder. EGTT is achieved by recombining a source's content with a reference's tone, while EGTR is attained by a novel tone masking objective that enforces content-tone disentanglement during training and realizes the removal procedure at inference. To improve transfer to tones unseen in training, a second training stage shapes a smooth tone space through variational sampling and audio-effects augmentation. Experimental results from both objective and subjective evaluations demonstrate that EG-VAE outperforms task-specific baselines on transfer and removal. Demos are available at \url{https://guitar-tone-demo.vercel.app/}.
\end{abstract}


\section{Introduction}
\label{introduction}

\begin{figure}[t]
\begin{center}
  \includegraphics[width=1.0\columnwidth]{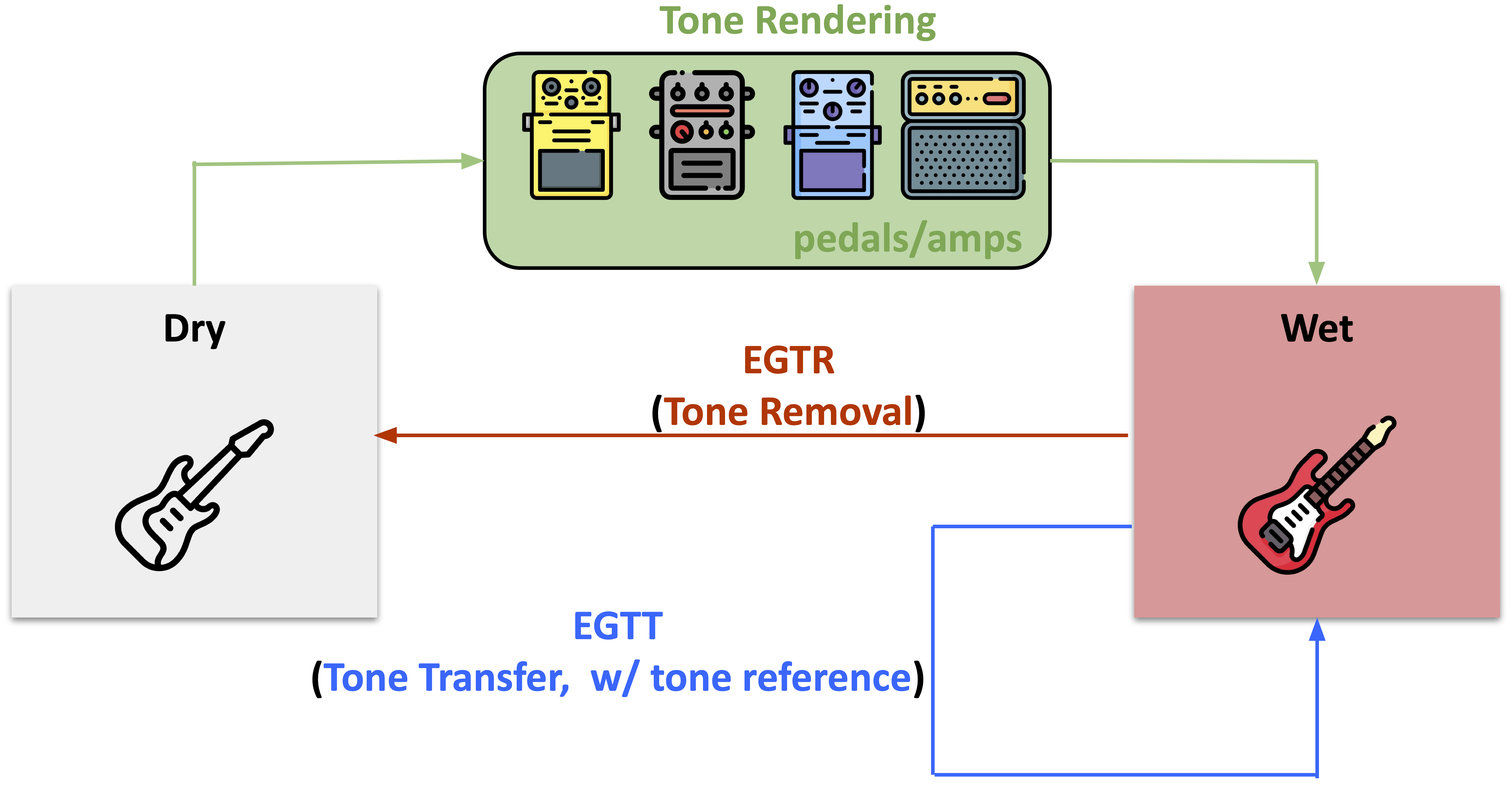}
  \caption{The two tone-modeling tasks studied in this work. A dry direct-input (DI) signal is rendered into a wet recording by a chain of effects pedals and an amplifier (top, green). Tone removal (EGTR, red) recovers the dry signal from the wet recording, inverting the rendering process. Tone transfer (EGTT, blue) replaces the tone of the wet recording with that of another wet reference, preserving the played content.}
  \label{fig:tone_intro}
\end{center}
\end{figure}

The electric guitar has been a long-standing subject of study in the music information retrieval (MIR) and audio effects communities, across tasks such as transcription~\cite{chen2022towards, chen2025towards, riley2024high, stein2010automatic, zang2024synthtab}, performance rendering~\cite{loth2025guitarflow}, and pedal/amplifier modeling~\cite{eichas2017block, wright2019real, wright2021neural, yeh2025ddsp}. In a typical electric guitar recording, the \textit{dry} direct-input (DI) signal is processed by a chain of effects pedals and an amplifier, producing the \textit{wet} signal heard by the listener~\cite{chen2022towards}. The character imparted by this signal chain is the recording's \textit{tone}, whose variation has direct consequences across these tasks. Transcription accuracy, for instance, degrades substantially under heavily processed tones~\cite{chen2025towards}. Despite this central role, tone remains challenging to model and control. 
In particular, we focus on the interplay between the following two tasks that operate on tone directly:
\textit{electric guitar tone transfer} (EGTT), which replaces a recording's tone with that of another, and \textit{electric guitar tone removal} (EGTR), which recovers the dry DI signal from a wet recording. Figure~\ref{fig:tone_intro} illustrates the two tasks and their relationship to the underlying signal chain.

Although singing voice~\cite{yu2025improving} and general music~\cite{steinmetz2024st, koo2025ito, steinmetz2022style} have shown promising results in audio effects style transfer, EGTT remains challenging~\cite{steinmetz2024st}, owing to the highly nonlinear distortion and the wide diversity of effects (e.g., modulation, equalization, and spatial effects; see  Section~\ref{sec:bg-signal-chain}) its signal chain. We observe that the common formulation of effects transfer is not well suited to EGTT. These methods estimate a tone from a reference and apply it to the input, implicitly assuming the input is \emph{dry} or only lightly processed. This assumption fails for guitar, whose input is a \emph{wet} recording carrying a chain of non-invertible effects. A natural route to faithful transfer is thus to first remove the input's own tone, making removal a prerequisite for transfer rather than a separate task. Prior approaches that sidestep this issue,  which assumes a dry DI input and targets only the amplifier, are restricted cases of EGTT, addressing neither the wet-input setting nor the full signal chain~\cite{chen2024towards}.

This recovery of the dry DI signal from a wet recording is precisely the task of EGTR. Existing removal methods, however, still struggle to produce high-quality audio. Prior work frames EGTR as signal enhancement~\cite{imort2022distortion, lee2024distortion, rice2023general} and addresses only restricted settings: synthetic, simplified effects such as hard clipping~\cite{imort2022distortion}, a narrow range of effect types that excludes reverb and modulation~\cite{lee2024distortion}, or general audio rather than guitar~\cite{rice2023general}. None removes the effects involved in a \emph{full} guitar chain, by which we mean coverage of all common guitar pedal and amplifier types. More fundamentally, each treats removal as a standalone problem, producing a representation specialized to EGTR and disconnected from EGTT.

We argue that the limitations of both tasks share a single cause: each is formulated in isolation, overlooking that the two act on the same attribute, the recording's tone. Transfer replaces this tone with another, while removal strips it away to recover the dry signal. The two are therefore complementary, manipulating one shared object in opposite directions. This naturally raises the question: can unifying EGTT and EGTR via representation learning overcome the limitations of task-specific methods?

We propose EG-VAE, as shown in Figure~\ref{fig:architecture}, which factorizes a wet recording into a content embedding and a tone embedding through a variational autoencoder. Three challenges remain. First, the factorization is non-trivial to obtain, as tone information tends to leak into the content embedding, a persistent failure mode in disentangled representation learning~\cite{luo2024posterior}. Second, EGTR has no reference to specify its target: it must recover the dry signal from the wet input alone. We address both with the proposed \textit{tone masking}, which removes the tone from the decoder's input, leaving the content embedding to reconstruct the signal alone. This single mechanism serves two roles: it defines EGTR, as the masked forward pass directly produces the dry DI signal; and it suppresses leakage, since the content embedding, forced to reconstruct the dry signal without the tone, must learn to exclude tone information. The third challenge is generalization: at the inference time, EGTT must handle reference tones unseen during the training time. We address this in a second training stage with two complementary mechanisms: variational sampling under KL regularization, which shapes a smooth tone space so that tones near the training distribution decode coherently, and audio-effects augmentation, which broadens that distribution to better cover unseen tones.

We evaluate EG-VAE against existing specialized EGTT and EGTR baselines in two settings: seen tones, whose signal chains appear in training, and unseen tones, which assess generalization to chains held out from training. A single unified model handles both tasks: EG-VAE reduces the Mel spectral distance of EGTT by $44\%$ over the strongest task-specific baseline (i.e., `One-to-many' with EGTR enhancement~\cite{chen2024towards}, $0.86$ vs.\ $1.53$ on seen tones) and improves over the best EGTR model (i.e., `Distortion Recovery'~\cite{lee2024distortion}, $1.10$ vs.\ $1.21$ Mel on seen tones, matching on unseen). An ablation study further confirms the contribution of each disentanglement mechanism and the second-stage design on unseen tones.

Our main contributions are summarized as follows:
\begin{itemize}
    \item We propose EG-VAE, to our best knowledge the first unified framework for EGTT and EGTR, learning a single factorization of a wet electric guitar recording into content and tone that supports both tasks.
    \item We introduce tone masking, a single objective shared by training and inference that defines the EGTR procedure while reinforcing content--tone disentanglement.
    \item We design a two-stage training strategy that smooths the tone space through variational sampling and audio-effects augmentation, improving transfer to unseen tones.
\end{itemize}

This paper is organized as follows. Section~\ref{sec:background} reviews the electric guitar signal chain and prior work on EGTT and EGTR. Section~\ref{sec:method} presents EG-VAE, covering its problem formulation, architecture, and two-stage training procedure. Section~\ref{sec:expsetup} describes the experimental setup, including the dataset, baselines, and evaluation protocol. Section~\ref{sec:results} reports objective and subjective results together with ablation studies. Section~\ref{sec::conclusion} concludes the paper.
\section{Background and Related Work}
\label{sec:background}

\subsection{The Electric Guitar Signal Chain}
\label{sec:bg-signal-chain}
An electric guitar produces sound through magnetic pickups that convert string vibration into an electrical signal, the \textit{direct-input (DI)} signal or \textit{dry signal}, denoted $\mathbf{x}_\text{dry} \in \mathbb{R}^{T}$ for a recording of $T$ samples. To produce the sound delivered to the listener, the dry signal is routed through a \textit{signal chain} consisting of one or more audio effects processors followed by an amplifier, yielding the \textit{wet signal} $\mathbf{x}_\text{wet} \in \mathbb{R}^{T}$:
\begin{equation}
    \mathbf{x}_\text{wet} = f(\mathbf{x}_\text{dry}) = f_\text{amp}(f_\text{fx}(\mathbf{x}_\text{dry})),
\end{equation}
where $f: \mathbb{R}^{T} \to \mathbb{R}^{T}$ is the transformation applied by the signal chain, $f_\text{fx}$ denotes the composition of effects processors preceding the amplifier, and $f_\text{amp}$ denotes the amplifier.\footnote{We adopt the conventional signal flow in which the effects sub-chain precedes the amplifier. In practice, some effects may be placed after the amplifier, reverberation, for instance, but we omit such cases from the notation for simplicity, as they do not affect the formulation that follows.} We refer to $f$ as the recording's \textit{tone}. The effects sub-chain $f_\text{fx}$ is itself a composition of $M$ processors applied in sequence,
\begin{equation}
    f_\text{fx} = f_M \circ f_{M-1} \circ \cdots \circ f_1,
\end{equation}
where each $f_i$ corresponds to a single audio effects processor and $f_1$ is applied first. We note that each $f_i$ may be time-varying and signal-dependent, with internal parameters omitted from the notation for clarity. Audio effects processors span a broad design space; commonly recognized categories include \textit{distortion} (non-linear, e.g., overdrive, fuzz), \textit{modulation} (time-varying, e.g., chorus, flanger), \textit{equalization} (linear spectral shaping), and \textit{spatial} (convolutional, e.g., reverb), though many processors combine these or fall outside any single category. The amplifier $f_\text{amp}$, in turn, is itself a multi-stage processor, typically comprising a preamplifier, tone stack, power amplifier, and speaker cabinet, each contributing its own non-linear or filtering behavior~\cite{yeh2025ddsp}. Even within a recognized category, different implementations produce perceptually distinct sounds owing to differences in circuit topology and analog non-linearities~\cite{5280324, pakarinen2009review}. Combined with the combinatorial structure of both $f_\text{fx}$ and $f_\text{amp}$, this variation makes guitar tone a complex, high-dimensional function.

\subsection{Electric Guitar Tone Transfer}
\label{sec:bg-egtt}
Direct work on EGTT is limited. The closest prior work treats EGTT as a one-to-many audio effects modeling problem. Chen \emph{et al.}~\cite{chen2024towards} use contrastive-pretrained tone embeddings for amplifier modeling, training a neural network on paired (dry, wet) data to emulate multiple amplifiers conditioned on a reference recording. In the notation of Section~\ref{sec:bg-signal-chain}, this targets $f_\text{amp}$, while assuming the input is the dry DI signal $\mathbf{x}_\text{dry}$ with no effects chain $f_\text{fx}$. The broader audio effects style transfer literature instead targets $f_\text{fx}$, and treats it more flexibly. Steinmetz \emph{et al.}~\cite{steinmetz2022style} use differentiable signal processing~\cite{engel2020ddsp} to estimate $f_\text{fx}$ parameters end-to-end, requiring the order and type of each $f_i$ to be specified before training (e.g., parametric equalizer $\rightarrow$ compressor in their work); this constrains the model to a fixed configuration and limits its flexibility on signal chains that do not match the predefined structure. Subsequent work~\cite{yu2025improving, koo2025ito} relaxes this training-time requirement through inference-time optimization, but the chain structure itself remains fixed. ST-ITO~\cite{steinmetz2024st} further generalizes this line by lifting the predefined-chain requirement, supporting arbitrary $f_\text{fx}$ at inference.

Yet even with this flexibility, a more fundamental limitation remains, concerning the input itself rather than the chain structure. These methods impose no formal restriction on the input signal, but they ultimately produce an output of the form
\begin{equation}
    \hat{\mathbf{x}} = \hat{f}_\text{fx}(\mathbf{x}_\text{base}),
    \label{eq:chain-est}
\end{equation}
where the estimated chain $\hat{f}_\text{fx}$ is applied to a base signal $\mathbf{x}_\text{base}$. This works well when $\mathbf{x}_\text{base}$ is dry or only lightly processed. In the setting we target, however, the input is usually a wet recording already, $\mathbf{x}_\text{base} = f(\mathbf{x}_\text{dry})$, carrying an unknown chain $f$. Applying a new chain on top then yields $\hat{f}_\text{fx} \circ f$: the reference tone is stacked onto the existing one rather than replacing it. Faithful transfer would first require removing $f$, precisely the tone-removal problem (EGTR), which these methods do not address. This difficulty is reflected in ST-ITO~\cite{steinmetz2024st}, whose authors report guitar tone matching as a particularly challenging case.

Across these methods, EGTT thus remains an open problem. Prior work addresses either $f_\text{amp}$~\cite{chen2024towards} or $f_\text{fx}$~\cite{steinmetz2022style, steinmetz2024st} alone, and no previous work has been demonstrated in the setting where the input already carries an unknown chain that has to be removed before a new tone can be applied.

\subsection{Electric Guitar Tone Removal}
\label{sec:bg-egtr}
While tone transfer replaces one tone with another, tone removal seeks to undo the signal chain entirely: given a wet recording $\mathbf{x}_\text{wet} = f(\mathbf{x}_\text{dry})$, the goal is to recover the dry signal,
\begin{equation}
    \hat{\mathbf{x}}_\text{dry} = f^{-1}(\mathbf{x}_\text{wet}) \approx \mathbf{x}_\text{dry},
    \label{eq:egtr}
\end{equation}
i.e., to invert the unknown chain $f$. This inversion is fundamentally ill-posed: $f$ is generally non-invertible, as nonlinear stages such as distortion clip and saturate the signal, discarding information that cannot be analytically restored. Removal must therefore be \emph{learned} from data rather than computed in closed form, and--unlike transfer, which derives its target tone from a reference recording--removal has no reference: the dry target must be produced from the wet input alone.

Existing work has approached EGTR along two lines, both framing removal as a signal enhancement or specialized source separation problem~\cite{imort2022distortion, lee2024distortion, rice2023general}. Guitar-specific approaches focus on distortion-related effects: Imort \emph{et al.}~\cite{imort2022distortion} formulate distortion removal as source separation, and Lee \emph{et al.}~\cite{lee2024distortion} use a two-stage Mel-denoiser and HiFi-GAN vocoder~\cite{kong2020hifi} to remove a wider set of effects. According to the notation of Section~\ref{sec:bg-signal-chain}, these methods invert $f_\text{amp}$ together with the components of $f_\text{fx}$ that exclude modulation and spatial processing. A separate line targets general-purpose effect removal: RemFX~\cite{rice2023general} covers a broader effect set including reverb and chorus, but uses architectures developed for general audio and is not specialized to the structure of $f_\text{amp}$ or to guitar-specific instances of $f_\text{fx}$.

Two limitations follow from this framing. First, no single method inverts the \emph{full} guitar chain: guitar-specific approaches omit the modulation and spatial components of $f_\text{fx}$, while methods that cover these are not specialized to guitar. Second, and more fundamentally, treating removal as a standalone enhancement task yields a representation usable only for removal, disconnected from transfer even though both operate on the same function $f$. The possibility of modeling EGTR and EGTT jointly  remains unexplored, despite their shared structure suggesting the two tasks could benefit each other.

\subsection{Transfer Tasks in the Audio Domain}
\label{sec:bg-adjacent}


Beyond the task-specific prior work discussed above, three transfer tasks in the audio domain share methodological similarities with our setting but differ from guitar tone modeling in structurally important ways.

\textbf{Voice conversion} factorizes speech into linguistic content and speaker identity~\cite{qian2019autovc, ju2024naturalspeech, zhang2025vevo, zhang2025vevo2}, modifying the latter while preserving the former. Its disentangled factor, speaker identity, is a property of the physical \emph{source}, the vocal tract and articulatory habits of the speaker, and is bounded by what a human vocal tract can produce. Guitar tone is structurally different: the DI signal is fixed, and tone arises from the \emph{post-source} signal chain, spanning the non-linear, time-varying, and convolutional transformations of an arbitrary chain.

\textbf{Instrument timbre transfer} factorizes audio into musical content and instrument identity, typically converting between distinct instrument categories such as violin and flute~\cite{lee2026diffusion, baoueb2024wavetransfer, engel2020ddsp}. The disentangled factor is the instrument label itself: each instrument is treated as a single point in timbre space, ignoring how the timbre of single instrument varies. However, for electric guitar, this variation is exactly what we aim to model: one guitar can produce a wide range of tones, determined not by the instrument body but by its signal chain. A categorical, between-instrument formulation cannot capture this, as it operates at a coarser level of granularity. 


These two tasks fall short of guitar tone modeling for distinct reasons: voice conversion's disentangled factor is a bounded source property rather than a post-source transformation; and instrument timbre transfer captures only categorical, between-instrument identity rather than continuous, within-instrument variation.

\subsection{Disentangled Sequential Autoencoders}
\label{sec:bg-disentangled}
A disentangled sequential autoencoder (DSAE) is a generative model that factorizes an observed time sequence into two latent variables capturing distinct aspects of the data~\cite{li2018disentangled}. Given an input sequence $\mathbf{x}_{1:T}$ of length $T$, a sequence of frame-level latent variables $\mathbf{z}_{1:T}$ captures local, time-varying features, while a latent vector $\mathbf{s}$ captures global, time-invariant factors of variation. Under the model parameters $\theta$, the joint distribution factorizes as
\begin{equation}
    p_\theta(\mathbf{x}_{1:T}, \mathbf{z}_{1:T}, \mathbf{s}) = \prod_{t=1}^{T} p_\theta(\mathbf{x}_t \mid \mathbf{s}, \mathbf{z}_t)\, p_\theta(\mathbf{z}_t)\, p_\theta(\mathbf{s}),
\end{equation}
where the priors on $\mathbf{z}_t$ and $\mathbf{s}$ are independent, encouraging disentanglement between the local and global factors. The model is typically learned within the variational autoencoder (VAE) framework~\cite{kingma2013auto} by maximizing the evidence lower bound on $\log p_\theta(\mathbf{x}_{1:T})$, with KL regularization controlling the information capacity of $\mathbf{s}$ as in $\beta$-VAE~\cite{betavae}.

This latent structure aligns naturally with the guitar signal chain of Section~\ref{sec:bg-signal-chain}: the frame-level $\mathbf{z}_{1:T}$ corresponds to the time-varying content of the dry  signal, while the global $\mathbf{s}$ corresponds to the time-invariant tone $f$. However, this alignment alone is not sufficient for our setting. First, prior independence does not guarantee posterior disentanglement: in practice, the global factor is often absorbed by the temporal mean of the frame-level latent, allowing the decoder to ignore $\mathbf{s}$ and leaking tone information into $\mathbf{z}_{1:T}$~\cite{luo2024posterior}. Second, the DSAE defines only a generative factorization; it provides no mechanism for tone removal, which requires a designated operating point of $\mathbf{s}$ that the decoder maps to the dry signal. These two gaps motivate the mechanisms introduced in Section~\ref{sec:method}.
\section{EG-VAE}
\label{sec:method}

\begin{figure*}[t]
  \centering
  \includegraphics[width=.95\linewidth,trim={0 0 20 0},clip]{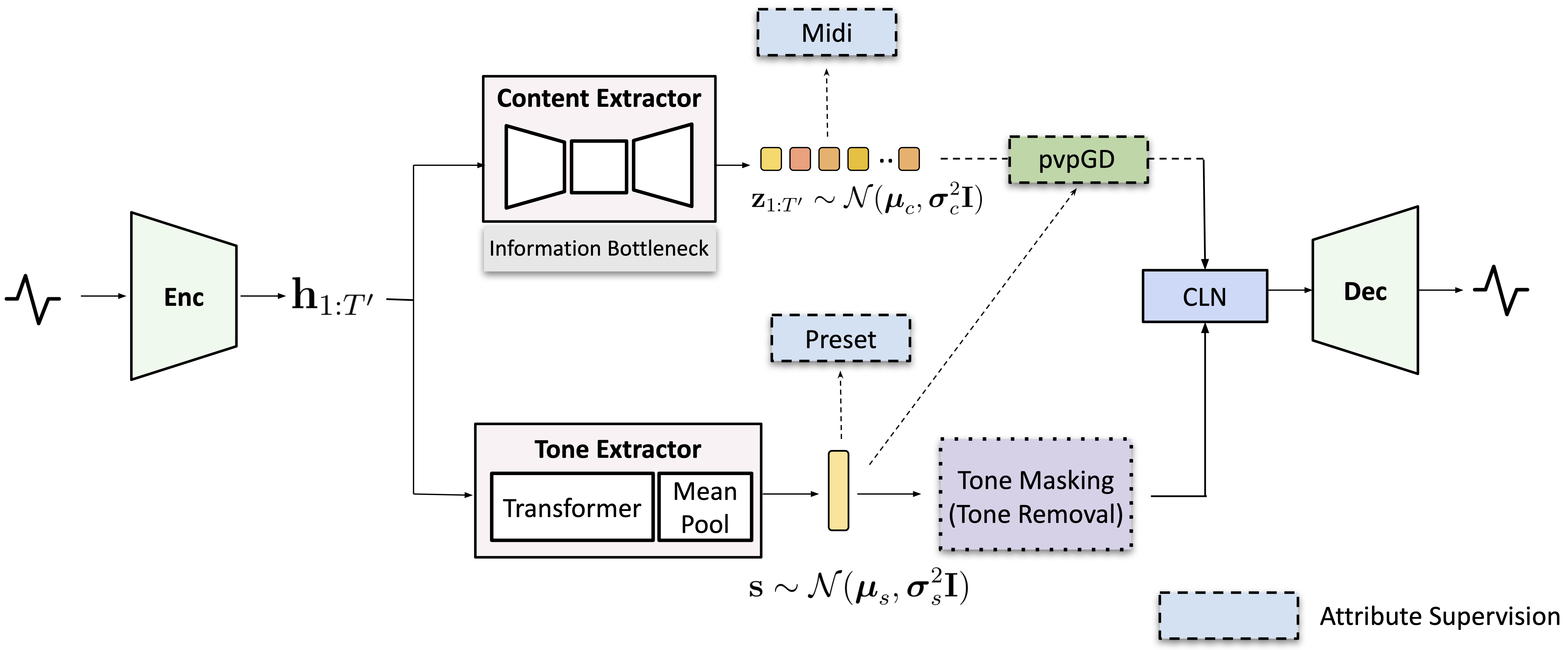}
  \vspace{3pt}
\caption{Overview of EG-VAE. A shared encoder maps the input waveform to features $\mathbf{h}_{1:T'}$, feeding a content extractor that produces a frame-level latent $\mathbf{z}_{1:T'} \sim \mathcal{N}(\boldsymbol{\mu}_c, \boldsymbol{\sigma}_c^2\mathbf{I})$ under an information bottleneck, and a tone extractor that produces a global tone embedding $\mathbf{s} \sim \mathcal{N}(\boldsymbol{\mu}_s, \boldsymbol{\sigma}_s^2\mathbf{I})$. The two are combined by conditional layer normalization (CLN) and decoded into the output. Dashed boxes are attribute-supervision heads; pvpGD perturbs $\mathbf{z}_{1:T'}$ to prevent collapse of $\mathbf{s}$. Tone masking overrides the CLN modulation ($\boldsymbol{\gamma}{=}\mathbf{1}, \boldsymbol{\beta}{=}\mathbf{0}$) and realizes tone removal. Sampling and pvpGD are active only in stage~2.}
\label{fig:architecture}
\end{figure*}

We present EG-VAE, a unified model for EGTT and EGTR. The model is trained in two stages: stage~1 establishes a disentangled factorization of wet recordings into content and tone embeddings, and stage~2 improves unseen tone generalizability by variational sampling on the tone embedding and audio-effects augmentation. 
We present the details below.

\subsection{Problem Formulation}
\label{sec:method-problem}

A wet recording $\mathbf{x}_\text{wet} \in \mathbb{R}^{T}$ is produced by applying a signal chain $f$ to a dry signal $\mathbf{x}_\text{dry} \in \mathbb{R}^{T}$.
We learn a representation that factorizes $\mathbf{x}_\text{wet}$ into two disentangled attributes: a frame-level \textit{content embedding} $\mathbf{z}_{1:T'}$ and a global \textit{tone embedding} $\mathbf{s}$. The content embedding $\mathbf{z}_{1:T'}$, of length $T' < T$, operates at a temporal resolution downsampled from the input by the encoder and captures the time-varying content of the dry signal, including the playing style; the tone embedding $\mathbf{s}$ captures the signal chain $f$. Ideally, $\mathbf{s}$ should encode the identity, ordering, and parameters of the audio effects processors composing $f$, while being invariant to the musical content. Therefore, the model consists of an encoder $\mathcal{E}$ producing $(\mathbf{z}_{1:T'}, \mathbf{s}) = \mathcal{E}(\mathbf{x}_\text{wet})$ and a decoder $\mathcal{D}$ producing $\hat{\mathbf{x}} = \mathcal{D}(\mathbf{z}_{1:T'}, \mathbf{s})$. 

Both EGTT and EGTR are  realized by manipulating the tone embedding $\mathbf{s}$ while preserving the content embedding. For EGTT, $\mathbf{s}$ is replaced with the tone embedding $\mathbf{s}^\text{ref}$ of a reference recording, yielding
\begin{equation}
    \hat{\mathbf{x}}^\text{EGTT} = \mathcal{D}(\mathbf{z}^\text{src}_{1:T'}, \mathbf{s}^\text{ref}),
\end{equation}
whose content and reference need not share musical content. For EGTR, the tone is instead removed rather than replaced, mapping the input to its dry signal, $\hat{\mathbf{x}}^\text{EGTR} \approx \mathbf{x}_\text{dry}$; we develop this masking operation in Section~\ref{sec:method-masking}.

\subsection{Model Architecture}
\label{sec:method-arch}

EG-VAE adapts the DAC-VAE~\cite{dacvae}, the VAE-version of DAC codec~\cite{kumar2023high},  for disentangled representation learning. As shown in Figure~\ref{fig:architecture}, a shared encoder maps the wet input to a frame-level latent, from which two extractors produce the embeddings $\mathbf{z}_{1:T'}$ and $\mathbf{s}$. A conditional layer normalization (CLN) module~\cite{chen2021adaspeech} then combines them, and a decoder reconstructs the audio waveform.

\paragraph{Shared encoder and decoder}
Rather than using two separate encoders for the two factors~\cite{luo2019learning, qian2019autovc}, we use a single shared encoder. Both content and tone are derived from the same underlying acoustic features, leaving the attribute-specific factorization to the two lightweight extractors that follow. The encoder is a stack of 1D convolutional residual blocks with Snake activations~\cite{ziyin2020neural} and progressive temporal downsampling, mapping the wet input $\mathbf{x}_\text{wet} \in \mathbb{R}^{T}$ to a frame-level latent $\mathbf{h} \in \mathbb{R}^{D \times T'}$, where $D$ is the channel dimension and $T'$ is the downsampled temporal resolution. The decoder mirrors this structure with progressive temporal upsampling, reconstructing the waveform from the CLN-combined representation $\tilde{\mathbf{z}}_{1:T'}$. 

\paragraph{Content and tone extractors}
The content extractor produces $\mathbf{z}_{1:T'}$ through a per-frame variational bottleneck. A $1\times1$ convolution projects $\mathbf{h}$ to per-frame posterior parameters $(\boldsymbol{\mu}_c, \boldsymbol{\sigma}_c)$ of a diagonal Gaussian, from which $\mathbf{z}_t \sim \mathcal{N}(\boldsymbol{\mu}_{c,t}, \boldsymbol{\sigma}_{c,t}^{2}\mathbf{I})$ is sampled; a second $1\times1$ convolution projects the sampled embedding back to the latent dimensionality. The tone extractor produces the global $\mathbf{s}$ in two steps: a transformer encoder processes $\mathbf{h}$ to capture dependencies across the temporal dimension, and a temporal pooling operation aggregates the result into a single fixed-dimensional vector. A linear variational head then produces the global posterior parameters $(\boldsymbol{\mu}_s, \boldsymbol{\sigma}_s)$, from which $\mathbf{s} \sim \mathcal{N}(\boldsymbol{\mu}_s, \boldsymbol{\sigma}_s^{2}\mathbf{I})$ is sampled.

\paragraph{Embedding combination via CLN}
The CLN module combines the two embeddings as
\begin{equation}
    \tilde{\mathbf{z}}_t = \boldsymbol{\gamma}(\mathbf{s}) \odot \text{LayerNorm}(\mathbf{z}_t) + \boldsymbol{\beta}(\mathbf{s}),
    \label{eq:cln}
\end{equation}
where $\boldsymbol{\gamma}(\mathbf{s})$ and $\boldsymbol{\beta}(\mathbf{s})$ are produced by a linear projection of $\mathbf{s}$, and $\text{LayerNorm}(\cdot)$~\cite{ba2016layer} omits learnable affine parameters. The combined $\tilde{\mathbf{z}}_{1:T'}$ is then decoded into the waveform.

\subsection{Disentanglement Mechanisms}
\label{sec:method-disentangle}
The architecture provides a factorization into $\mathbf{z}_{1:T'}$ and $\mathbf{s}$, but does not guarantee disentanglement directly. The core difficulty is that the frame-level latent $\mathbf{z}_{1:T'}$ may be expressive enough to encode the global tone factor across its frames, letting the decoder reconstruct the wet signal while ignoring $\mathbf{s}$. Our design therefore limits the capacity of $\mathbf{z}_{1:T'}$ so that the global factor is forced into $\mathbf{s}$. We apply four mechanisms during training: an information bottleneck on $\mathbf{z}_{1:T'}$ limits its per-frame capacity; content-tone perturbation forces each embedding to encode only its intended factor by recombining content and tone drawn from different recordings; attribute supervision provides external targets for what each embedding should encode; and posterior variance-parameterised Gaussian dropout (pvpGD)~\cite{luo2024posterior} prevents posterior collapse of the tone latent. A fifth mechanism, tone masking, is the key to unifying the two tasks and is introduced separately in Section~\ref{sec:method-masking}.

\paragraph{Information bottleneck on $\mathbf{z}_{1:T'}$}
Inspired by~\cite{qian2019autovc}, we apply a KL regularization on $\mathbf{z}_{1:T'}$ against a standard normal prior to constrain its capacity:
\begin{equation}
    \mathcal{L}_\text{KL}^c = \mathrm{KL}\!\big(q(\mathbf{z}_{1:T'} \mid \mathbf{x}_\text{wet}) \,\|\, p(\mathbf{z})\big), \quad p(\mathbf{z}) = \mathcal{N}(\mathbf{0}, \mathbf{I}).
    \label{eq:klc}
\end{equation}
The regularization is applied in a lower-dimensional space: the encoder features are first projected from $1024$ to $128$ dimensions, where the KL term is computed, and then projected back to the original dimensionality to form the content embedding. Weighted by $\lambda_\text{KL}^c$, this term limits the information capacity of $\mathbf{z}_{1:T'}$, as in $\beta$-VAE~\cite{betavae}, pressuring it to encode only the information necessary for reconstruction.

\paragraph{Content-tone perturbation}
Inspired by~\cite{choi2021neural}, this mechanism forces each embedding to be invariant to the factor it should not encode. Let $\mathbf{x}_\text{wet}^{(c, t)}$ denote a wet recording with content identifier $c$ (the dry  signal it is rendered from) and tone identifier $t$ (the preset applied to that dry signal). At each training step we sample a content-match $\mathbf{x}_\text{wet}^{(c_0, t_1)}$, a tone-match $\mathbf{x}_\text{wet}^{(c_1, t_0)}$, and a target $\mathbf{x}_\text{wet}^{(c_0, t_0)}$. We extract the content embedding from the content-match, the tone embedding from the tone-match, and supervise the decoder to reconstruct the target:
\begin{equation}
    \mathcal{L}_\text{rec}^{\text{c\text{-}t}} = \mathcal{L}_\text{audio}\!\big(\mathcal{D}(\mathbf{z}_{1:T'}^{(c_0, t_1)}, \mathbf{s}^{(c_1, t_0)}),\ \mathbf{x}_\text{wet}^{(c_0, t_0)}\big).
    \label{eq:perturb}
\end{equation}
Neither input matches the target on both factors, since the content-match supplies content $c_0$ with a different tone $t_1$, while the tone-match supplies tone $t_0$ with a different content $c_1$. Reconstructing $\mathbf{x}_\text{wet}^{(c_0, t_0)}$ therefore requires $\mathbf{z}_{1:T'}^{(c_0, t_1)}$ to encode only $c_0$ and $\mathbf{s}^{(c_1, t_0)}$ to encode only $t_0$; any residual leakage in either direction hurts the reconstruction.

\paragraph{Attribute supervision}
Inspired by~\cite{ju2024naturalspeech}, we supervise each embedding with an auxiliary classification task on the attribute it should capture. The content embedding $\mathbf{z}_{1:T'}$ is passed to a frame-level pitch classifier that predicts, for each frame, the set of MIDI pitches active at that frame. Because guitar performance is polyphonic, multiple pitches may be active simultaneously, so this is formulated as multi-label classification with binary cross-entropy applied independently per pitch ($\mathcal{L}_\text{pitch}$). The tone embedding $\mathbf{s}$ is passed to a global \emph{preset} classifier, where a preset---a specific combination of effects and amplifier settings with their parameter values---is the standard unit for describing a guitar tone. Our training data is annotated with one preset label per recording, which we use as a discrete tone identifier to guide $\mathbf{s}$; this classifier is trained with categorical cross-entropy ($\mathcal{L}_\text{preset}$). However, because preset supervision is categorical, it promotes separation between training tones but does not guarantee a continuous, smooth tone space; we address this in stage~2 (Section~\ref{sec:method-training}).

\paragraph{pvpGD (probabilistic perturbation guided by tone variance)}
Another disentanglement difficulty, distinct from the leakage above, is posterior collapse of the global latent: under KL regularization on $\mathbf{s}$, the posterior variance can grow toward $\boldsymbol{\sigma}_s \to \mathbf{1}$, at which point $\mathbf{s}$ loses informativeness and the frame-level latent $\mathbf{z}_{1:T'}$ absorbs the modeling capacity that should reside in $\mathbf{s}$~\cite{luo2024posterior}. This collapses the factorization: tone information migrates back into $\mathbf{z}_{1:T'}$, the precise leakage our other mechanisms work to prevent. To avoid this, we adopt the pvpGD mechanism of~\cite{luo2024posterior}, which couples the capacity of $\mathbf{z}_{1:T'}$ to that of $\mathbf{s}$. Before CLN combination, the sampled frame-level latent is perturbed by multiplicative Gaussian noise whose scale is set by the tone posterior variance:
\begin{equation}
    \mathbf{z}_t^\text{pvp} = \mathbf{z}_t \odot \big(1 + \bar{\sigma}_s \, \boldsymbol{\epsilon}_t\big), \quad \boldsymbol{\epsilon}_t \sim \mathcal{N}(\mathbf{0}, \mathbf{I}),
    \label{eq:pvpgd}
\end{equation}
where $\bar{\sigma}_s = \exp\!\big(\tfrac{1}{D_s}\sum_{i=1}^{D_s} \log \sigma_{s,i}\big)$ is the geometric mean of the per-dimension tone-posterior standard deviations, with $D_s$ the tone-embedding dimensionality and $\sigma_{s,i}$ the $i$-th component of $\boldsymbol{\sigma}_s$, and $\mathbf{z}_t^\text{pvp}$ replaces $\mathbf{z}_t$ as the input to CLN in Eq.~\eqref{eq:cln}. This couples the two latents from both sides: when $\boldsymbol{\sigma}_s$ grows—the signature of collapse—the perturbation on $\mathbf{z}_{1:T'}$ strengthens and degrades reconstruction, penalizing collapse; conversely, shrinking $\boldsymbol{\sigma}_s$ to reclaim capacity for $\mathbf{z}_{1:T'}$ is penalized by $\mathcal{L}_\text{KL}^s$. The two pressures bound $\boldsymbol{\sigma}_s$, holding $\mathbf{s}$ at a capacity that retains tone information. pvpGD is thus a disentanglement mechanism in nature: it keeps the tone latent informative so the factorization does not collapse. Because it requires a stochastic tone path, however, it is operative only once variational sampling is introduced in stage~2 (Section~\ref{sec:method-training}); we therefore evaluate its effect in the stage-2 ablation (Table~\ref{tab:ablation-training}). At inference, $\bar{\sigma}_s$ is set to zero and the perturbation vanishes.

\subsection{Tone Masking}
\label{sec:method-masking}
\begin{figure}[t]
\begin{center}
  \includegraphics[width=1.0\columnwidth]{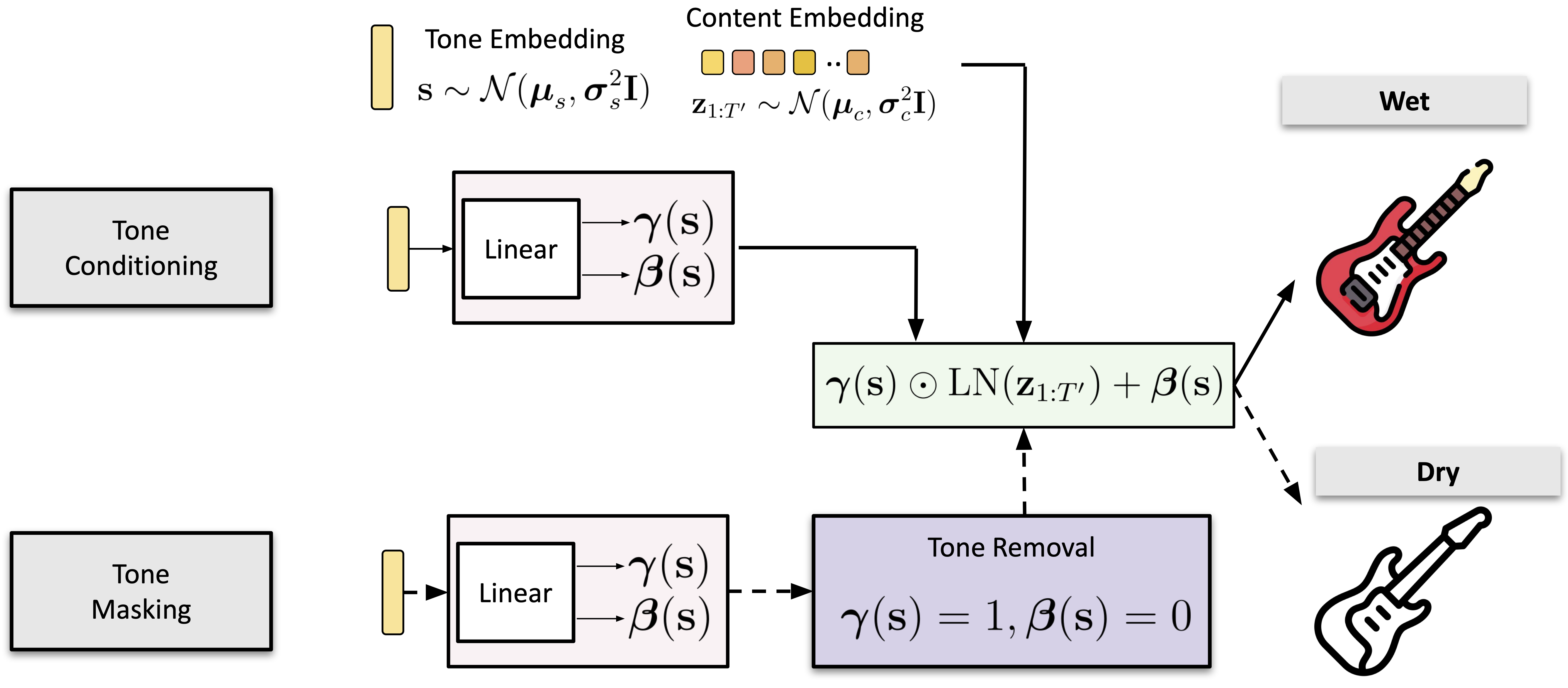}
  \caption{Tone conditioning versus tone masking, sharing the same CLN operation $\boldsymbol{\gamma}(\mathbf{s}) \odot \text{LN}(\mathbf{z}_{1:T'}) + \boldsymbol{\beta}(\mathbf{s})$. Conditioning produces $\boldsymbol{\gamma}(\mathbf{s}), \boldsymbol{\beta}(\mathbf{s})$ from the tone embedding, yielding a wet output; masking overrides them to $\boldsymbol{\gamma}=\mathbf{1}, \boldsymbol{\beta}=\mathbf{0}$, leaving the content unmodulated and yielding the dry DI signal. Tone removal is thus the masked decoding path.}
  \label{fig:tone_mask}
\end{center}
\end{figure}

Tone masking removes the tone from the combined representation, leaving the content embedding to reconstruct the dry  signal alone (Figure~\ref{fig:tone_mask}). This serves two roles: a disentanglement objective during the training time, and a direct way to execute EGTR at  the inference time.

\paragraph{The masking forward pass}
Recall that CLN modulates the content embedding through $\boldsymbol{\gamma}(\mathbf{s})$ and $\boldsymbol{\beta}(\mathbf{s})$ (Eq.~\eqref{eq:cln}). Masking overrides this modulation directly, setting $\boldsymbol{\gamma} = \mathbf{1}$ and $\boldsymbol{\beta} = \mathbf{0}$ regardless of $\mathbf{s}$, so the content embedding passes through unmodulated:
\begin{equation}
    \tilde{\mathbf{z}}_t = \text{LayerNorm}(\mathbf{z}_t).
\end{equation}
We call this the \textit{no-modulation regime} and write the corresponding decoding as $\mathcal{D}_\text{mask}(\mathbf{z}_{1:T'})$. Because the override acts directly on $\boldsymbol{\gamma}$ and $\boldsymbol{\beta}$, it is independent of the value of $\mathbf{s}$ and of how the CLN projection is trained.

This operation has a direct interpretation in terms of normalization. Layer normalization removes a representation's global statistics, and conditional normalization layers re-introduce a global factor through their affine parameters $(\boldsymbol{\gamma}, \boldsymbol{\beta})$, a mechanism widely used for style transfer~\cite{huang2017arbitrary} and attribute-conditioned synthesis~\cite{chen2021adaspeech}. CLN follows this design, injecting the tone factor through $(\boldsymbol{\gamma}(\mathbf{s}), \boldsymbol{\beta}(\mathbf{s}))$. Setting $\boldsymbol{\gamma} = \mathbf{1}$ and $\boldsymbol{\beta} = \mathbf{0}$ thus reduces CLN to plain layer normalization, and tone masking acts as the inverse of tone conditioning.

\paragraph{Training (masking as a disentanglement objective)}
At training, the masked pass is supervised against the dry  signal:
\begin{equation}
    \mathcal{L}_\text{mask} = \mathcal{L}_\text{audio}\!\big(\mathcal{D}_\text{mask}(\mathbf{z}_{1:T'}),\ \mathbf{x}_\text{dry}\big).
    \label{eq:mask}
\end{equation}
Under the no-modulation regime, $\mathbf{z}_{1:T'}$ is the decoder's only input, so it must carry everything needed to reconstruct $\mathbf{x}_\text{dry}$. Together with the bottleneck $\mathcal{L}_\text{KL}^c$, which caps its capacity, this leaves no room for tone-related information that the dry target does not require, improving the content-tone factorization. The objective also fixes the meaning of the no-modulation regime as the state the decoder maps to the dry signal, consistent with the dry targets in our data (Section~\ref{sec:method-problem}).

\paragraph{Inference (masking as the EGTR procedure)}
EGTR reuses this masked pass unchanged, $\hat{\mathbf{x}}^\text{EGTR} = \mathcal{D}_\text{mask}(\mathbf{z}_{1:T'})$. Because training and inference share an identical operation, EGTR needs no extra module, fine-tuning, or task-specific procedure: it is a direct readout of the representation learned for EGTT. Accordingly, EG-VAE unifies the two tasks.

\subsection{Unseen Tone Generalizability}
\label{sec:method-smoothness}
At inference, EGTT must handle \textit{unseen} reference tones, those extracted from recordings whose signal chains are not in the training set. 
To improve generalization, the following two properties are desirable: 1) the decoder varies \textit{smoothly} with $\mathbf{s}$, so that a tone near a training tone decodes coherently; 2) the training tones \textit{cover} the space densely enough such that unseen tones fall near them. 
They correspond to how the decoder responds to $\mathbf{s}$, and how the training tones are distributed.
We address these by two mechanisms in stage~2 training.

\paragraph{Variational sampling on the tone path}
For the decoder to vary smoothly with $\mathbf{s}$, we treat $\mathbf{s}$ as a stochastic latent and sample $\mathbf{s} \sim \mathcal{N}(\boldsymbol{\mu}_s, \boldsymbol{\sigma}_s^2\mathbf{I})$ during training, rather than using a deterministic, unregularized timbre embedding as~\cite{ju2024naturalspeech, liu2026synthcloner}. Each step exposes the decoder to a neighborhood of $\boldsymbol{\mu}_s$, forcing it to produce coherent output across that neighborhood rather than at isolated points~\cite{kingma2013auto}; the KL regularization $\mathcal{L}_\text{KL}^s$ extends this continuity globally by pulling the posterior toward a standard normal prior. A deterministic embedding offers no such guarantee, supporting neither sampling of new tones nor smooth interpolation~\cite{stanton2022speaker, hsu2018hierarchical}.

\paragraph{Audio-effects augmentation}
We augment each preset's recordings with additional effects, producing variants that sit near the preset but realize slightly different chain configurations. This densifies the distribution of training tones, so that an unseen tone is more likely to fall near data the decoder has seen. Rather than arbitrary effects, the augmentation follows the Wiener--Hammerstein model, a linear--nonlinear--linear cascade widely used to model guitar amplifiers~\cite{yeh2025ddsp}: gain and parametric equalization form the linear stages and distortion the nonlinearity. Applying the distortion with a fixed probability lets the chain realize either a linear or a nonlinear system, covering both regimes of guitar signal chains. 

\subsection{Training Procedure}
\label{sec:method-training}
The complete training objective combines audio reconstruction, KL, and attribute supervision terms:
\begin{equation}
    \mathcal{L}_\text{total} = \mathcal{L}_\text{audio} + \lambda_\text{KL}^c \mathcal{L}_\text{KL}^c + \lambda_\text{KL}^s \mathcal{L}_\text{KL}^s + \lambda_\text{pitch} \mathcal{L}_\text{pitch} + \lambda_\text{preset} \mathcal{L}_\text{preset},
\end{equation}
where $\mathcal{L}_\text{audio} = \lambda_\text{mel} \mathcal{L}_\text{mel} + \lambda_\text{adv} \mathcal{L}_\text{adv} + \lambda_\text{fm} \mathcal{L}_\text{fm}$ combines a multi-scale mel-spectrogram loss with the adversarial and $L_1$ feature-matching losses against a jointly trained discriminator. The tone masking objective $\mathcal{L}_\text{mask}$ is realized through $\mathcal{L}_\text{audio}$ in the masked-reconstruction mode. Loss weights and other hyperparameters are given in Section~\ref{sec:exp-impl}.

\paragraph{Training modes.}
Each step samples one of three modes, all extracting content from a \textit{content-match} recording $\mathbf{x}_\text{wet}^{(c_0, \cdot)}$ and differing in the tone reference and target. \textit{Self-reconstruction} draws the tone from $\mathbf{x}_\text{wet}^{(c_1, t_0)}$ and targets $\mathbf{x}_\text{wet}^{(c_0, t_0)}$. \textit{Conversion} draws the tone from an external recording $\mathbf{x}_\text{wet}^{(c_2, t_2)}$ and targets $\mathbf{x}_\text{wet}^{(c_0, t_2)}$, exposing the model to tone references beyond the batch. \textit{Masked reconstruction} masks the tone and targets the dry signal $\mathbf{x}_\text{dry}$, realizing $\mathcal{L}_\text{mask}$.

\paragraph{Two-stage training}
Stage~1 establishes the disentangled factorization under simplified conditions: the tone path is deterministic (no variational head, $\lambda_\text{KL}^s = 0$), pvpGD is off, and no data augmentation. Mode sampling, $\mathcal{L}_\text{KL}^c$, $\mathcal{L}_\text{pitch}$, and $\mathcal{L}_\text{preset}$ jointly enforce disentanglement and ground the masked state to the dry signal. We omit the variational head here because sampling $\mathbf{s}$ from random initialization destabilizes the disentanglement objectives and hinders convergence. Stage~2 initializes from the stage-1 checkpoint and adds the smoothness components (Section~\ref{sec:method-smoothness}): the variational head produces $(\boldsymbol{\mu}_s, \boldsymbol{\sigma}_s)$ for sampling $\mathbf{s}$, $\mathcal{L}_\text{KL}^s$ is warmed up to ease the transition from the deterministic path, pvpGD prevents collapse of the stochastic tone latent, and audio-effects augmentation broadens the tone manifold. Preset classification is disabled ($\lambda_\text{preset} = 0$), since augmentation breaks the correspondence between each recording and its preset label.

\paragraph{Inference}
The encoder is deterministic, using $(\boldsymbol{\mu}_c, \boldsymbol{\mu}_s)$ and disabling pvpGD. EGTT combines the content embedding of a source recording with the tone embedding of a reference; EGTR applies the masked forward pass.
\section{Experimental Setup}
\label{sec:expsetup}

\subsection{Datasets}
Because no existing electric guitar dataset offers realistic recordings with sufficient tonal diversity, we construct our own using commercial audio plugins from \emph{Neural DSP}.\footnote{\url{https://neuraldsp.com/}} We take the dry DI signals from the EGDB dataset~\cite{chen2022towards} and render them through two plugins: \texttt{Archetype:\,Cory~Wong~X}\footnote{\url{https://neuraldsp.com/plugins/archetype-cory-wong}} provides the \textit{seen} tones for training and validation, and \texttt{Morgan~Amps~Suite}\footnote{\url{https://neuraldsp.com/plugins/morgan-amps-suite}} provides the \textit{unseen} tones used exclusively for evaluation. The two plugins implement different effects and amplifiers, so the evaluation tones are produced by signal chains unseen during training.

Each plugin exposes a complete guitar signal chain spanning all four effect categories mentioned in Section~\ref{sec:bg-signal-chain}: for instance, \texttt{Archetype:\,Cory~Wong~X} includes distortion (overdrive, booster, and tube amplifiers with cabinet simulation), modulation (chorus and an envelope filter), equalization (a 9-band graphic EQ), and spatial effects (delay and reverb). We render each plugin's official factory presets, $218$ for \texttt{Archetype:\,Cory~Wong~X} and $96$ for \texttt{Morgan~Amps~Suite}, each a professionally designed tone exercising a different configuration of this chain. To further increase seen-tone diversity, we additionally include the EGDB-PG dataset~\cite{chen2025towards}, which provides $256$ amplifier-only tones.

The EGDB DI signals amount to roughly $2$ hours of mono audio, which we first split into training, validation, and test partitions in an $80/5/15$ ratio. Each split is then rendered through its corresponding tone set, ensuring no DI content overlaps between training and evaluation. The training partition is rendered through the $256 + 218 = 474$ seen tones, yielding approximately $1.6 \times 474 \approx 758$ hours of training audio. The test partition supports two evaluation settings: rendering it through the $474$ seen tones yields approximately $0.3 \times 474 \approx 142$ hours for seen-tone evaluation, and rendering it through the $96$ \texttt{Morgan~Amps~Suite} presets yields $0.3 \times 96 \approx 29$ hours for unseen-tone evaluation over signal chains held out from training. Compared with existing electric guitar datasets, ours offers substantially greater tonal diversity and broader coverage of the signal chain: EGDB~\cite{chen2022towards} provides only $5$ presets, EGDB-PG~\cite{chen2025towards} $256$, and SynthTab~\cite{zang2024synthtab} $23$, and all three cover amplifiers alone, rather than the full pedal-and-amplifier chain considered here.

\subsection{Implementation Details}
\label{sec:exp-impl}
EG-VAE builds on the DAC-VAE backbone. The encoder uses downsampling rates $[2, 8, 10, 12]$ for a hop length of $1920$, and the decoder mirrors it with rates $[12, 10, 8, 2]$, operating at a $44.1$\,kHz sample rate. The content extractor applies its variational bottleneck at dimension $128$ before projecting back to the $1024$-dimensional latent consumed by the decoder; the tone embedding has dimension $64$. The tone extractor is a $4$-layer feed-forward Transformer with $4$ attention heads, followed by mean pooling and the variational head.

We follow the discriminator and adversarial setup of DAC-VAE: a multi-period waveform discriminator with periods $\{2, 3, 5, 7, 11\}$ and a multi-scale complex STFT discriminator with FFT sizes $\{2048, 1024, 512\}$ and band splits at $\{0, 0.1, 0.25, 0.5, 0.75, 1.0\}$ of the Nyquist frequency. The multi-scale mel loss uses $n_\text{mels} \in \{5, 10, 20, 40, 80, 160, 320\}$ with corresponding window lengths $\{32, 64, \dots, 2048\}$, and the multi-scale STFT loss uses window lengths $\{2048, 512\}$.

Both generator and discriminator are optimized with AdamW ($\beta_1 = 0.8$, $\beta_2 = 0.99$, learning rate $1.5\times10^{-4}$) and an exponential learning-rate decay ($\gamma = 0.9999996$). Models are trained on $1$-second segments with a batch size of $14$. Stage~1 trains for $150{,}000$ steps; stage~2 fine-tunes from the stage-1 checkpoint for $200{,}000$ steps, with the tone KL weight $\lambda_\text{KL}^s$ linearly warmed up over the first $10{,}000$ steps. The three training modes (self-reconstruction, conversion, masked reconstruction) are sampled with probabilities $[0.25, 0.45, 0.30]$. We use loss weights $\lambda_\text{mel} = 15$, $\lambda_\text{fm} = 2$, $\lambda_\text{adv} = 1$, $\lambda_\text{KL}^c = \lambda_\text{KL}^s = 10^{-4}$, and (in stage~1) $\lambda_\text{pitch} = 1$, $\lambda_\text{preset} = 5$. The stage-2 audio-effects augmentation (Section~\ref{sec:method-smoothness}) is implemented with the \texttt{audiomentations} library\footnote{\url{https://github.com/iver56/audiomentations}} and applied to each preset recording with probability $0.5$. All configurations are trained on a single NVIDIA RTX PRO 6000 GPU. For each configuration, we select the checkpoint with the lowest validation loss for evaluation.

\subsection{Baselines}
\label{sec:exp-baselines}
\paragraph{EGTT baselines.}
We compare against two existing models: 
One-to-many~\cite{chen2024towards}, the amplifier-modeling approach, and DeepAFx~\cite{steinmetz2022style}, a differentiable audio-effects style-transfer method, using its strongest configuration with a chain spanning all effect categories: gain $\rightarrow$ parametric EQ $\rightarrow$ compressor $\rightarrow$ distortion $\rightarrow$ reverberation. 
Neither is designed for wet input: 
One-to-many  assumes a dry signal, while DeepAFx places no explicit restriction. 
To give them the most favorable comparison,
we train each baseline in two configurations. The standard configuration takes the wet recording directly as input. The \textit{EGTR-enhanced} configuration (denoted ``\emph{w/ EGTR}'') is trained on dry input and, at inference, sources from a dry signal recovered by EG-VAE's tone removal. Since EG-VAE attains the best removal performance among all methods (cf. Table~\ref{tab:egtr}), this gives the baselines the strongest available dry estimate, relieving them of the wet-input mismatch (Section~\ref{sec:bg-egtt}).

\paragraph{EGTR baselines.}
We compare against four tone-removal baselines spanning the two lines discussed in Section~\ref{sec:bg-egtr}. HDemucs~\cite{defossez2021hybrid}, DCUNet~\cite{choi2018phase}, and DPTNet~\cite{chen2020dual} are source-separation and enhancement architectures adapted to recover the dry DI signal, and Distortion Recovery~\cite{lee2024distortion} is a specialized model for guitar effect removal. We follow the baseline selection of prior removal work~\cite{rice2023general, lee2024distortion}.

\paragraph{Training setup.}
All baselines are retrained on the same data as EG-VAE, following the architecture and training setup of each original work, with the optimization objective standardized to the multi-scale mel-spectrogram loss used in DAC-VAE for a consistent comparison across methods. For Distortion Recovery~\cite{lee2024distortion}, we additionally replace the original HiFi-GAN vocoder with the more recent BigVGAN~\cite{lee2022bigvgan} for higher-fidelity synthesis.

\subsection{Evaluation Protocol}
\label{sec:exp-protocol}
We evaluate EG-VAE along five axes: tone transfer (EGTT, Table~\ref{tab:egtt}), tone removal (EGTR, Table~\ref{tab:egtr}), tone-space smoothness (Table~\ref{tab:smooth_ppl}), an ablation study over its components (Tables~\ref{tab:ablation-disentangle} and~\ref{tab:ablation-training}), and a qualitative analysis of the learned representation. Unless otherwise noted, experiments report results on both \textit{seen} and \textit{unseen} tones, with all objective metrics averaged over $2{,}000$ ten-second test examples.

\paragraph{Objective metrics}
Following~\cite{kumar2023high}, we report two spectral distances between the model output and the task target: the \textit{Mel distance}, the $L_1$ distance between log-mel spectrograms, and the \textit{STFT distance}, the $L_1$ distance between log-magnitude spectrograms, which better captures fidelity at higher frequencies. The target is the reference-tone recording for EGTT, and the dry DI signal for EGTR, respectively. Lower values indicate closer agreement.

\paragraph{Subjective metrics}
As spectral distances do not fully reflect perceptual quality, we also conduct a listening test with $14$ participants, rating each example on a $1$--$5$ scale. For both tasks, listeners rate \textit{audio quality} (AQ), the overall fidelity and naturalness of the output, independent of the task objective. Each task adds a task-specific criterion: \textit{tone similarity} for EGTT, measuring how closely the output's tone matches the reference tone, and \textit{dryness} for EGTR, measuring how closely the output resembles a clean DI signal, i.e., how completely the signal-chain processing has been removed. We additionally include the ground-truth target as an upper-bound anchor (denoted as \emph{Oracle} in Tables~\ref{tab:egtt} and~\ref{tab:egtr}), providing a reference for the highest rating listeners assign for each criterion.

\paragraph{Tone-space smoothness}
To assess whether the tone space is smooth, we adopt the perceptual path length (PPL)~\cite{karras2019style}, which measures the perceptual change in the decoded output under small displacements along an interpolation path in the embedding space. Concretely, we interpolate between the tone embeddings of two reference recordings while holding the content fixed; at $32$ random positions $t$ along the path, we measure the distance between outputs decoded at $t$ and $t+\epsilon$. We evaluate over $500$ such interpolation pairs, yielding $500 \times 32 = 16{,}000$ samples. Following~\cite{karras2019style}, we normalize each distance by $\epsilon^2$ (with $\epsilon = 10^{-4}$); as the perceptual distance we use the $L_1$ distance between normalized log-mel spectrograms. A lower PPL indicates that small steps in tone space produce small output changes, i.e., a smoother space. We report PPL on unseen tones, the setting smoothness is designed to support.

\begin{table*}
\centering
\caption{Tone transfer results on seen and unseen tones. Mel and STFT are objective spectral distances, while AQ (Audio Quality) and TS (Tone Similarity) are subjective ratings from the listening test.}
\label{tab:egtt}
\small
\setlength{\tabcolsep}{4pt}
\begin{tabular}{l | c c c c | c c c c}
\toprule
\multirow{2}{*}{Model} & \multicolumn{4}{c|}{Seen Tone / \texttt{Archetype: Cory Wong X}} & \multicolumn{4}{c}{Unseen Tone / \texttt{Morgan Amps Suite}} \\
\cmidrule(lr){2-5} \cmidrule(lr){6-9}
& Mel $\downarrow$ & STFT $\downarrow$ & MOS (AQ) $\uparrow$ & MOS (TS) $\uparrow$ & Mel $\downarrow$ & STFT $\downarrow$ & MOS (AQ) $\uparrow$ & MOS (TS) $\uparrow$ \\
\midrule
Oracle                    & --- & --- & 4.26 $\pm$ 0.71 & 4.48 $\pm$ 0.64 & --- & --- & 3.85 $\pm$ 1.06 & 4.48 $\pm$ 0.70 \\
\midrule
DeepAFx~\cite{steinmetz2022style}                   & 2.07 $\pm$ 0.74 & 3.0 $\pm$ 0.84 & 2.85 $\pm$ 1.23 & 1.44 $\pm$ 0.58 & 1.63 $\pm$ 0.49 & 2.58 $\pm$ 0.57 & 2.19 $\pm$ 0.83 & 1.81 $\pm$ 0.83 \\
\quad w/ EGTR            & 2.04 $\pm$ 0.64 & 2.98 $\pm$ 0.73 & 3.37 $\pm$ 1.08 & 2.30 $\pm$ 1.23 & 1.75 $\pm$ 0.43 & 2.72 $\pm$ 0.49 & 3.15 $\pm$ 0.91 & 2.74 $\pm$ 1.20 \\
One-to-many~\cite{chen2024towards}               & 1.93 $\pm$ 0.62 & 2.85 $\pm$ 0.83 & 3.67 $\pm$ 1.11 & 2.30 $\pm$ 0.82 & 1.69 $\pm$ 0.44 & 2.7 $\pm$ 0.64 & 3.67 $\pm$ 0.78 & 2.63 $\pm$ 0.93 \\
\quad w/ EGTR             & 1.53 $\pm$ 0.42 & 2.55 $\pm$ 0.83 & 3.11 $\pm$ 0.89 & 2.32 $\pm$ 0.99 & 1.61 $\pm$ 0.35 & 2.61 $\pm$ 0.64 & 2.56 $\pm$ 0.97 & 2.07 $\pm$ 0.78 \\
\midrule
EG-VAE                    & \textbf{0.86 $\pm$ 0.17} & \textbf{1.63 $\pm$ 0.15} & \textbf{4.15 $\pm$ 0.66} & \textbf{4.30 $\pm$ 0.61} & \textbf{1.15 $\pm$ 0.24} & \textbf{2.00 $\pm$ 0.28} & \textbf{3.85 $\pm$ 0.82} & \textbf{3.70 $\pm$ 0.72} \\
\bottomrule
\end{tabular}
\end{table*}

\paragraph{Ablation study}
We ablate EG-VAE in two groups, reflecting the two stages of training. The first (Table~\ref{tab:ablation-disentangle}) removes individual disentanglement mechanisms active in stage~1: MIDI (pitch) supervision, tone (preset) classification, content--tone perturbation, and tone masking, and evaluates each variant across three modes: reconstruction, tone transfer, and tone removal. As these mechanisms govern the factorization itself rather than generalization, we evaluate this group on seen tones only. Removing tone masking (\emph{w/o Tone Masking}) disables EGTR entirely, so its removal column is left empty. The second group (Table~\ref{tab:ablation-training}) ablates the stage-2 components against the full model: variational sampling, audio-effects augmentation, and pvpGD. We note that pvpGD appears here rather than in the first group because, as a mechanism guarding against posterior collapse, it is only operative once the variational tone path is introduced in stage~2. 

\paragraph{Representation visualization}
To assess the factorization qualitatively, we visualize tone embeddings on seen and unseen reference tones with t-SNE~\cite{van2008visualizing}. A well-disentangled representation should yield tone embeddings that cluster by preset, capturing tone identity.
\section{Results and Discussion}
\label{sec:results}

\begin{table*}
\centering
\caption{Tone removal results on seen and unseen tones. Mel and STFT are objective spectral distances, while AQ (Audio Quality) and Dry (Dryness) are subjective ratings from the listening test.}
\label{tab:egtr}
\small
\setlength{\tabcolsep}{4pt}
\begin{tabular}{l | c c c c | c c c c}
\toprule
\multirow{2}{*}{Model} & \multicolumn{4}{c|}{Seen Tone / \texttt{Archetype: Cory Wong X}} & \multicolumn{4}{c}{Unseen Tone / \texttt{Morgan Amps Suite}} \\
\cmidrule(lr){2-5} \cmidrule(lr){6-9}
& Mel $\downarrow$ & STFT $\downarrow$ & MOS\,(AQ)\,$\uparrow$ & MOS\,(Dry)\,$\uparrow$ & Mel $\downarrow$ & STFT $\downarrow$ & MOS\,(AQ)\,$\uparrow$ & MOS\,(Dry)\,$\uparrow$ \\
\midrule
Oracle                    & --- & --- & 4.67 $\pm$ 0.63 & 4.58 $\pm$ 0.65 & --- & --- & 4.50 $\pm$ 0.74 & 4.39 $\pm$ 0.77 \\
\midrule
HDemucs~\cite{defossez2021hybrid}                 & 1.36 $\pm$ 0.39 & 2.48 $\pm$ 0.89 & 3.00 $\pm$ 1.31 & 2.69 $\pm$ 1.17 & 1.63 $\pm$ 0.39 & 2.63 $\pm$ 0.88 & 3.03 $\pm$ 1.06 & 2.39 $\pm$ 1.15 \\
DCUNet~\cite{choi2018phase}                   & 1.54 $\pm$ 0.47 & 2.77 $\pm$ 0.91 & 3.36 $\pm$ 1.05 & 2.83 $\pm$ 0.97 & 1.57 $\pm$ 0.43 & 2.71 $\pm$ 0.85 & 3.19 $\pm$ 1.01 & 2.56 $\pm$ 1.11 \\
DPTNet~\cite{chen2020dual}                    & 1.19 $\pm$ 0.40 & 2.01 $\pm$ 0.61 & 2.81 $\pm$ 1.12 & 2.19 $\pm$ 0.95 & 1.27 $\pm$ 0.37 & 2.06 $\pm$ 0.63 & 2.67 $\pm$ 1.33 & 2.31 $\pm$ 1.28 \\
Distortion Recov.~\cite{lee2024distortion}       & 1.21 $\pm$ 0.34 & 1.85 $\pm$ 0.61 & 1.53 $\pm$ 0.61 & 2.36 $\pm$ 1.13 & 1.22 $\pm$ 0.34 & \textbf{1.86 $\pm$ 0.62} & 2.06 $\pm$ 0.92 & 2.11 $\pm$ 1.09 \\
\midrule
EG-VAE                    & \textbf{1.10 $\pm$ 0.24} & \textbf{1.79 $\pm$ 0.43} & \textbf{3.53 $\pm$ 0.84} & \textbf{3.42 $\pm$ 0.69} & \textbf{1.19 $\pm$ 0.28} & 1.87 $\pm$ 0.50 & \textbf{3.69 $\pm$ 0.86} & \textbf{3.50 $\pm$ 0.74} \\
\bottomrule
\end{tabular}
\end{table*}

\subsection{Tone Transfer (EGTT)}

Table~\ref{tab:egtt} reports the tone transfer results. EG-VAE achieves the lowest spectral distances by a clear margin, reaching $0.86$ Mel and $1.63$ STFT on seen tones versus $1.53$ and $2.55$ for the strongest baseline (One-to-many w/ EGTR), and maintaining its lead on unseen tones ($1.15$ Mel, $2.00$ STFT) despite a slight drop relative to seen tones.

The EGTR-enhanced baselines confirm our analysis of the wet-input setting (Section~\ref{sec:bg-egtt}): supplying a tone-removed input improves both methods on seen tones (One-to-many $1.93 \to 1.53$ Mel, DeepAFx $2.07 \to 2.04$), yet both remain well behind EG-VAE. We attribute this gap to the two-stage nature of the baselines, each difficult in its own right. DeepAFx relies on a differentiable chain of fixed type and order, which cannot model chains deviating from its structure, while One-to-many's more flexible gated convolutional backbone still does not capture the full diversity of guitar tones. A two-stage pipeline thus compounds removal and modeling errors, whereas EG-VAE transfers on a single representation, avoiding such error accumulation.

Notably, the weaker baselines score lower on unseen than on seen tones, the opposite of EG-VAE. This reflects their weak fit in both settings rather than better generalization: having fit the seen tones poorly, they have no seen-tone advantage to lose, whereas EG-VAE's seen-best ordering is the expected signature of a model that fits the trained tones and degrades gracefully on held-out ones.

Subjective ratings show similar trends with objective results. EG-VAE achieves the highest AQ and TS (tone similarity) on both seen ($4.15$, $4.30$) and unseen ($3.85$, $3.70$) tones, approaching the Oracle reference ($4.26$/$4.48$ seen, $3.85$/$4.48$ unseen) and demonstrating strong generalization. On AQ, all baselines remain reasonably close to EG-VAE---unsurprisingly, since effects-transfer methods fundamentally preserve audio fidelity. The main gap lies in TS (tone similarity), where the baselines cluster at low scores ($1.81$--$2.74$) with no consistent preference among them, and EG-VAE nearly doubles the next-best score on both settings. The gap widens further on unseen tones: even One-to-many w/ EGTR, the objectively strongest baseline on seen tones, drops to $2.07$, reflecting the difficulty for amplifier-targeted methods of applying effects beyond their training distribution. In contrast, EG-VAE's smaller drop ($4.30 \to 3.70$) reflects the smooth tone space shaped in stage~2 (Section~\ref{sec:results-smoothness}).

\subsection{Tone Removal (EGTR)}
Table~\ref{tab:egtr} reports tone removal results. On seen tones, EG-VAE attains the best spectral distances ($1.10$ Mel, $1.79$ STFT), outperforming all removal baselines. On unseen tones, it remains strongest in Mel distance ($1.19$) and is comparable to Distortion Recovery in STFT ($1.87$ versus $1.86$). This shows the two tasks can be unified without sacrificing either.

\begin{figure}[t]
\begin{center}
  \includegraphics[width=0.8\columnwidth]{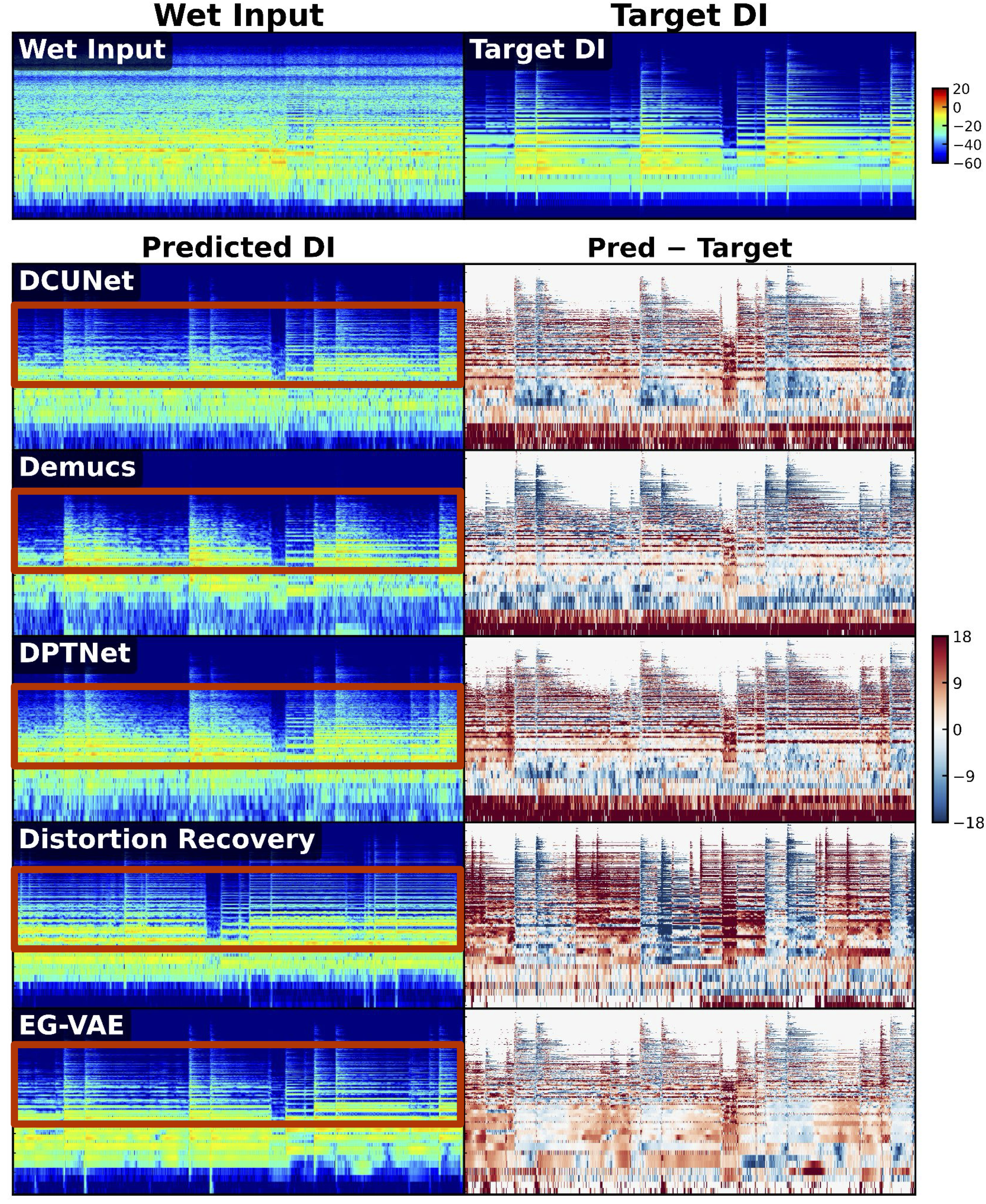}
  \caption{Tone removal on a representative distortion-heavy tone. Top: wet input and target DI. For each method: predicted DI (left, log-magnitude dB) and the residual between prediction and target.}
  \label{fig:egtr}
\end{center}
\end{figure}

\begin{table*}
\centering
\caption{Ablation study results across three evaluation modes. Values are reported as mean $\pm$ std.}
\label{tab:ablation-disentangle}
\small
\setlength{\tabcolsep}{4pt}
\begin{tabular}{l | c c | c c | c c}
\toprule
\multirow{2}{*}{Model} & \multicolumn{2}{c|}{Reconstruction} & \multicolumn{2}{c|}{Tone Transfer} & \multicolumn{2}{c}{Tone Removal} \\
\cmidrule(lr){2-3} \cmidrule(lr){4-5} \cmidrule(lr){6-7}
& Mel $\downarrow$ & STFT $\downarrow$ & Mel $\downarrow$ & STFT $\downarrow$ & Mel $\downarrow$ & STFT $\downarrow$ \\
\midrule
EG-VAE            & \textbf{0.80 $\pm$ 0.12} & \textbf{1.58 $\pm$ 0.13} & \textbf{0.86 $\pm$ 0.16} & \textbf{1.64 $\pm$ 0.16} & \textbf{1.13 $\pm$ 0.27} & \textbf{1.84 $\pm$ 0.49} \\
\midrule
\multicolumn{7}{l}{\textit{Ablation Studies}} \\
\quad w/o MIDI & 1.02 $\pm$ 0.18 & 1.81 $\pm$ 0.16 & 1.10 $\pm$ 0.25 & 1.89 $\pm$ 0.21 & 1.56 $\pm$ 0.30 & 2.37 $\pm$ 0.48 \\
\quad w/o Tone Classification & 0.85 $\pm$ 0.15 & 1.64 $\pm$ 0.15 & 0.95 $\pm$ 0.23 & 1.73 $\pm$ 0.21 & 1.21 $\pm$ 0.28 & 1.97 $\pm$ 0.51 \\
\quad w/o Perturbation & 1.48 $\pm$ 0.39 & 2.57 $\pm$ 0.44 & 1.67 $\pm$ 0.51 & 2.74 $\pm$ 0.56 & 2.12 $\pm$ 0.56 & 3.15 $\pm$ 1.17 \\
\quad w/o Tone Masking & 0.96 $\pm$ 0.18 & 1.76 $\pm$ 0.15 & 1.02 $\pm$ 0.23 & 1.83 $\pm$ 0.20 & - & - \\
\bottomrule
\end{tabular}
\end{table*}

\begin{table*}
\centering
\caption{Ablation study results across three evaluation modes.}
\label{tab:ablation-training}
\small
\begin{tabular}{l | c c | c c | c c}
\toprule
\multirow{2}{*}{Model} & \multicolumn{2}{c|}{Reconstruction} & \multicolumn{2}{c|}{Tone Transfer} & \multicolumn{2}{c}{Tone Removal} \\
\cmidrule(lr){2-3} \cmidrule(lr){4-5} \cmidrule(lr){6-7}
& Mel $\downarrow$ & STFT $\downarrow$ & Mel $\downarrow$ & STFT $\downarrow$ & Mel $\downarrow$ & STFT $\downarrow$ \\
\midrule
EG-VAE (two stage) & 1.02 $\pm$ 0.15 & \textbf{1.88 $\pm$ 0.24} & \textbf{1.14 $\pm$ 0.21} & \textbf{2.00 $\pm$ 0.28} & \textbf{1.18 $\pm$ 0.27} & \textbf{1.86 $\pm$ 0.49} \\
\midrule
\multicolumn{7}{l}{\textit{Ablation Studies}} \\
\quad w/o Variational Sampling & 1.10 $\pm$ 0.17 & 1.96 $\pm$ 0.27 & 1.25 $\pm$ 0.23 & 2.10 $\pm$ 0.31 & 1.24 $\pm$ 0.30 & 1.94 $\pm$ 0.54 \\
\quad w/o Fx-Aug & 1.10 $\pm$ 0.16 & 1.99 $\pm$ 0.27 & 1.23 $\pm$ 0.23 & 2.10 $\pm$ 0.31 & 1.25 $\pm$ 0.31 & 1.95 $\pm$ 0.58 \\
\quad w/o PvpGD  & \textbf{1.01 $\pm$ 0.16} & 1.89 $\pm$ 0.28 & 1.15 $\pm$ 0.23 & 2.02 $\pm$ 0.32 & 1.20 $\pm$ 0.30 & 1.88 $\pm$ 0.54 \\
\bottomrule
\end{tabular}
\end{table*}

Figure~\ref{fig:egtr} illustrates this on a distortion-heavy example. The difficulty lies in the harmonic region marked by the red boxes: nonlinear distortion introduces dense high-order harmonics absent from the dry DI. In this region, the baselines leave large residuals (right column), either over-suppressing or failing to remove the added harmonics, whereas EG-VAE's residual is visibly smaller. This reflects the ill-posedness of inverting nonlinear distortion (Section~\ref{sec:bg-egtr}): the added harmonics are hard to separate from the underlying signal, and EG-VAE's content representation, trained to reconstruct the dry DI under tone masking, captures it more accurately than the enhancement-based baselines.

The subjective results reveal a notable divergence from the objective ranking. While Distortion Recovery achieves competitive spectral scores (Mel $1.21$, STFT $1.85$ on seen), its perceived quality is the lowest among all methods ($1.53$ AQ on seen, $2.06$ on unseen), indicating that minimizing spectral distance alone does not yield a perceptually clean DI signal. In contrast, EG-VAE leads on both AQ ($3.53$ seen, $3.69$ unseen) and `Dry' (dryness) ($3.42$ seen, $3.50$ unseen), approaching the Oracle ($4.58$, $4.39$ dryness) and remaining stable across seen and unseen tones---suggesting that joint training under tone masking produces outputs listeners consistently judge as both faithful and dry.

\subsection{Ablation Study}
\paragraph{Disentanglement mechanisms}
Table~\ref{tab:ablation-disentangle} ablates the disentanglement mechanisms across the reconstruction, tone transfer, and tone-removal modes. Every mechanism contributes, removing any one degrades performance, confirming that each plays a distinct role in the factorization. Content--tone perturbation is the most apparent: removing it degrades all three modes most severely (reconstruction Mel $0.80 \to 1.48$, tone removal $1.13 \to 2.12$), as this strategy is what forces each embedding to retain only its intended factor; without it, content and tone are not factorized into the desired embedding. Removing MIDI supervision also harms all modes substantially (reconstruction Mel $0.80 \to 1.02$, tone removal $1.13 \to 1.56$). Distortion obscures the harmonic structure of the played notes in the spectrogram, making the content hard to recover; supervising $\mathbf{z}_{1:T'}$ with per-frame pitch counteracts this by explicitly preserving the playing content. Removing preset classification has a smaller effect on reconstruction and tone transfer but still degrades tone removal ($1.13 \to 1.21$ Mel), consistent with its role in shaping the tone embedding rather than the content path.

Removing tone masking disables tone removal entirely (shown as ``--''), since masking is the mechanism by means of which removal is performed. It also degrades reconstruction and tone transfer ($0.80 \to 0.96$ Mel reconstruction), showing that masking is not merely a removal procedure but also strengthens the content--tone factorization.

\paragraph{Stage-2 components}
Table~\ref{tab:ablation-training} ablates the stage-2 components. Removing variational sampling degrades most modes, most notably tone transfer (Mel $1.14 \to 1.25$, STFT $2.00 \to 2.10$) and tone-removal STFT ($1.86 \to 1.94$). The KL regularization $\mathcal{L}^s_\text{KL}$ organizes the tone embeddings into a continuous, Gaussian-structured manifold, so that tones near the training distribution, including unseen ones, decode coherently. A deterministic embedding lacks this structure and generalizes worse to held-out chains. Removing audio-effects augmentation similarly degrades tone transfer and removal (e.g., removal Mel $1.18 \to 1.25$), consistent with its role in broadening the tone space toward unseen configurations. Removing pvpGD produces a small but consistent degradation in tone transfer, removal, and smoothness (Tables~\ref{tab:ablation-training} and \ref{tab:smooth_ppl}) at a negligible reconstruction cost: by coupling the two latents to prevent collapse of $\mathbf{s}$, it keeps the tone embedding informative, trading marginal reconstruction fidelity for a more usable tone representation.

\begin{table}
\centering
\caption{Tone-space smoothness on unseen tones. Lower is smoother.}
\label{tab:smooth_ppl}
\small
\begin{tabular}{l | c}
\toprule
Model & Smoothness (PPL) $\downarrow$ \\
\midrule
EG-VAE     & \textbf{0.90} \\
\midrule
\multicolumn{2}{l}{\textit{Ablation Studies}} \\
\quad w/o Variational Sampling & 2.00  \\
\quad w/o Fx-Aug & 0.91  \\
\quad w/o PvpGD  & 0.96   \\
\bottomrule
\end{tabular}
\end{table}

\subsection{Tone-Space Smoothness}
\label{sec:results-smoothness}
Table~\ref{tab:smooth_ppl} reports the perceptual path length (PPL) on unseen tones, measuring how smoothly the decoded output varies along interpolation paths in the tone space; lower values indicate a smoother space. We see that EG-VAE achieves the smoothest space ($0.90$), and removing any stage-2 component increases PPL. The effect is by far largest for variational sampling ($0.90 \to 2.00$), confirming it as the primary driver of smoothness: without the stochastic, KL-regularized tone path, interpolating between two tones produces abrupt, less coherent changes in the output. This matches the design rationale of Section~\ref{sec:method-smoothness}, where sampling $\mathbf{s}$ under KL regularization is the mechanism that makes the decoder vary continuously with the tone embedding.

Removing either audio-effects augmentation or pvpGD increases PPL only marginally ($0.91$ and $0.96$ versus $0.90$). Both contribute to smoothness, but far less than variational sampling, which is consistent with their distinct roles: augmentation broadens the region covered by the tone space, and pvpGD keeps $\mathbf{s}$ informative by preventing collapse, whereas variational sampling shapes the continuous structure on which these two operate.

\subsection{Representation Visualization}
\label{sec:results-viz}
\begin{figure}[t]
\begin{center}
  \includegraphics[width=1.0\columnwidth]{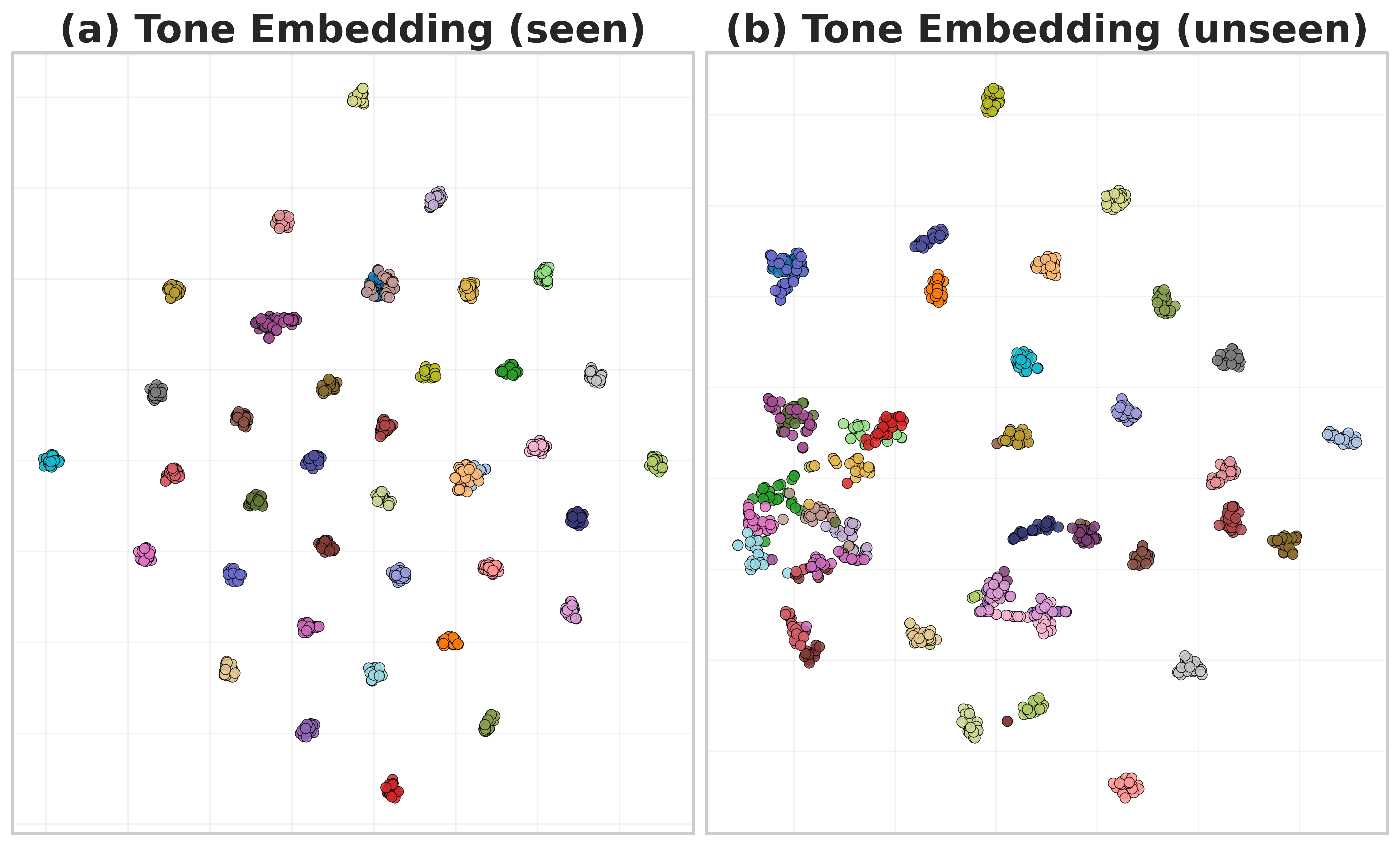}
  \caption{t-SNE visualization of the tone embedding $\mathbf{s}$, colored by preset, for (a) seen tones and (b) unseen tones.}
  \label{fig:tsne_tone}
\end{center}
\end{figure}

To qualitatively examine the tone representation, we visualize the tone embeddings $\mathbf{s}$ with t-SNE in Fig.~\ref{fig:tsne_tone}, coloring each point by its preset label. For seen tones (Fig.~\ref{fig:tsne_tone}a), the embeddings form clear, well-separated clusters: recordings sharing a signal chain map to nearby points while different presets occupy distinct regions, indicating that $\mathbf{s}$ captures tone identity. For unseen tones (Fig.~\ref{fig:tsne_tone}b), the clusters remain identifiable but less separated, consistent with the seen--unseen performance gap in transfer and removal. 

\section{Conclusion}
\label{sec::conclusion}
We presented EG-VAE, a unified framework for electric guitar tone transfer (EGTT) and tone removal (EGTR). Rather than treating the two tasks separately, EG-VAE learns a single disentangled representation of wet recordings, factorizing them into a content embedding and a tone embedding and realizing both tasks. Transfer recombines content with a reference tone, while removal is realized by a novel tone masking objective that plays a dual role: it strengthens content--tone disentanglement during training and defines the removal procedure at inference, using one operation identical in both. A second training stage shapes a smooth tone space, improving transfer quality on tones unseen during training. 

Several directions remain for future work. The content embedding could be factorized further into the played notes (e.g., MIDI) and the playing performance, enabling independent control over what is played and how. A harder challenge is delay: a delayed repetition is physically an effect yet acoustically resembles performed content, giving the model no consistent cue for assigning it to tone or content, an open problem for tone disentanglement. Finally, extending the framework to other instruments would broaden its applicability.

\bibliographystyle{IEEEtran}
\bibliography{main}

\vfill

\end{document}